\documentclass[journal=jpcl,manuscript=article]{achemso}
\usepackage[version=3]{mhchem}

\usepackage{graphicx} 
\usepackage{subfig}
\usepackage{multirow}
\usepackage{longtable}
\usepackage{achemso} 
\usepackage{array,booktabs}
\usepackage{float}
\newcolumntype{C}[1]{>{\centering\arraybackslash}p{#1}}
\usepackage{caption}
\usepackage{color}
\usepackage[]{siunitx}
\usepackage{verbatim} 
\setkeys{acs}{usetitle = true}
\usepackage{color, colortbl}
\definecolor{Gray}{gray}{0.9}
\usepackage{comment}
\usepackage{multirow}

\author{Anushna Bhattacharya$^\dag$}
\author{Vikas Tiwari$^\dag$}
\author{Tarak Karmakar}
\affiliation{Department of Chemistry, Indian Institute of Technology, Delhi,\\ Hauz Khas, New Delhi 110016, India}
\email{tkarmakar@chemistry.iitd.ac.in}

\title[\texttt{achemso}]
{Electrostatic-driven Self-assembly of Janus-like Monolayer-protected Metal Nanoclusters}

\begin{document}

\onecolumn
%\newpage
\maketitle
\begin{abstract}

{The generation of controlled microstructures of functionalized nanoparticles has been one of the crucial challenges in nanoscience and nanotechnology. Efforts have been made to tune ligand charge states that can affect the aggregation propensity and modulate the self-assembled structures. In this work, we modelled zwitterionic Janus-like monolayer ligand-protected metal nanoclusters (J-MPCs) and studied their self-assembly using atomistic molecular dynamics and advanced enhanced sampling simulations. The oppositely-charged ligands functionalization on two hemispheres of a J-MPC elicits asymmetric solvation, primarily driven by distinctive hydrogen bonding patterns in the ligand-solvent interactions. Electrostatic interactions between the oppositely charged residues in J-MPCs guide the formation of one-dimensional and ring-like self-assembled superstructures with molecular dipoles oriented in specific patterns. The pertinent atomistic insights into the intermolecular interactions governing the self-assembled structures of zwitterionic J-MPCs obtained from this work can be used to design a general strategy to create tunable microstructures of charged MPCs.}
\end{abstract}

\newpage
\section{INTRODUCTION}
Colloidal particles with functional constituents, concerning chemical functionalization, mostly have isotropic surfaces and homogeneous bulk properties. The self-assembly of these particles has not put forward much room to maneuver concerning the structural and behavioral properties\cite{Miszta2011,Schreiber2013,Grschel2013}. Hence, various synthetic methods have evolved, designing anatomically anisotropic particles\cite{Li2018}. Delving with the shape and functional group control at nanometer dimensions is primarily demanding. So, particles possessing anisotropic allocations of functional groups or structures have been synthesized. Named after the double-faced Roman god, \textit{Janus} particles are spherically-shaped particles having two completely distinct profiles on each hemisphere\cite{Walther2008}. These hemispheres might constitute different materials, and or bear different functional groups in the case of functionalized nanoparticles. The twofold identity of Janus particles results in these particles forming aggregates of various shapes and sizes having unique properties\cite{Lattuada2011}. The amphiphilic nature of their morphology is one of their most notable features, which causes them to behave as grainy surfactants, as investigated by Muller \textit{et al.}\cite{Kietzke2007,Wang2018,Walther2008,Walther2008_1,Grschel2012} Their ability to agglomerate at the interfaces helped us distinguish between a colloidal and a particulate surfactant. Self-assembly of magnetic Janus beads was investigated by Hatton \textit{et al.}, who showed the reversible self-assembled constitution of these beads by changing the pH\cite{Lattuada2007}. Drug delivery\cite{Tan2023, Tiwari2023biomat}, biological inquiry\cite{Su2019}, and virtual device manipulation\cite{Kang2018} are a few of the uses of these controlled self-assembled structures.

Monolayer-protected atomically-precise metal nanoclusters (MPCs), an extension of the nanocluster family much smaller than nanoparticles, are ligand-protected nanoclusters with a noble metal core surrounded by a monolayer of ligands\cite{Jin2016,Chakraborty2017}. Being atomically precise, these MPCs aid us in discerning its matter properties along with shape and size dependence at the atomic level\cite{Wilcoxon2006}. These nanoclusters offer great scope for us to regulate the interactions between one another to strategize their hierarchical self-assembly\cite{Bi2023}. Molecular level understanding of the self-assembly mechanism and optimization of factors modulating the aggregation process is of paramount importance for the design of 2D and 3D microstructures having potential applications. More importantly, controlling the self-assembly process and generating specific patterns by introducing directionality in the nanoclusters are challenging endeavors and have recently drawn the attention of the community\cite{Kolay2022}.

In this work, we have modeled zwitterionic J-MPCs with a metal core consisting of 102 Au atoms, which is protected by 22 negatively charged thiol-linked para-mercapto benzoate (\textit{p}MBA) ligands on one hemisphere and 22 positively charged thiol-linked para-aminothiophenol (\textit{p}ATP) ligands on the other and is represented by the formula Au$_{102}$(S-\textit{p}C$_6$H$_4$COO$^-$)$_{22}$(S-\textit{p}C$_6$H$_4$NH$_3^+$)$_{22}$. Molecular dynamics (MD) simulations of a single J-MPC in water revealed asymmetric solvation of the two sides of the particle. By inducing directionality through anisotropy, MPCs, as Janus particles (J-MPCs), are seen to have diverse interaction patterns and lead to various self-assembled structures such as linear, branched chains, and hexagonal rings with rich phase diagrams. We discuss here the mechanism of emergence of various patterns, characterize the thermodynamics of the self-assembly process, and discuss the role of solvent in the formation of various self-assembled structures.

\section{COMPUTATIONAL METHODS}

The monomer of Au$_{102}$(S-\textit{p}C$_6$H$_4$COO$^-$)$_{22}$(S-\textit{p}C$_6$H$_4$NH$_3^+$)$_{22}$ was simulated in a cubic box of volume around 118 nm$^3$ filled with water. The MPC multimers in water were simulated similarly, varying the box volume. Avogadro software\cite{Hanwell_2012}was used to create the initial structure of a pMBA ligand and pATP ligand. The NanoModeler server\cite{FrancoUlloa2019,Franco_Ulloa_2023} was used to create the full MPC structure and topology. It employs the bonded and non-bonded parameters of the gold-sulfur motifs developed by Pohjolainen et al.\cite{Pohjolainen2016} and Heinz et al.\cite{Heinz2008} The General AMBER Force Field\cite{Wang2004} was used for ligands, whereas we used the TIP3P potential model to describe the water molecules. All force-field parameters are provided in the supporting information (SI).

Using the steepest descent algorithm, the systems were first minimized, followed by thermally equilibrating them at a temperature of 300K in the canonical (NVT) ensemble. Thereafter, isothermal-isobaric (NPT) simulations were performed. The NVT and NPT MD simulations were carried out for 250 ps each having a time step of 0.5 fs, with the help of a Leap-Frog integrator. Using the stochastic velocity rescaling thermostat\cite{Bussi2007} with a temperature coupling constant of 0.5 ps, both NVT and NPT equilibriations were successfully executed. The pressure of the systems was maintained at 1 bar coupled with equilibriating the volumes using isotropic Parrinello-Rahman barostat.\cite{Parrinello1981} The Particle Mesh Ewald (PME) method with a grid spacing of 0.16 and an order of 4 was used to quantify the long-range electrostatic interactions. A cut-off of 1.0 nm was used for both the van der Waals and short-range Coulombic interactions. The partial-Mesh-Ewald method was used to treat the long-range electrostatic interactions. For the production NPT simulations, each system was simulated for a period of a few microseconds with a time step of 2 fs. The LINCS\cite{Hess1997} algorithm was used to constrain all covalent bonds involving hydrogen atoms during the production runs. GROMACS version 2021.4.6\cite{Berendsen_1995} was employed to perform all the simulations. The VMD software\cite{Humphrey1996} was used to visualize simulation trajectories and prepare figures and movies. A few GROMACS in-built tools along with in-house written scripts were used to perform the analyses.

\newpage
\noindent

\vspace{30pt}
\noindent
     
\section{RESULTS \& DISCUSSION}

\subsection{Monomer in water}

\begin{figure}[!h]
    \centering
    \includegraphics[width=165mm]{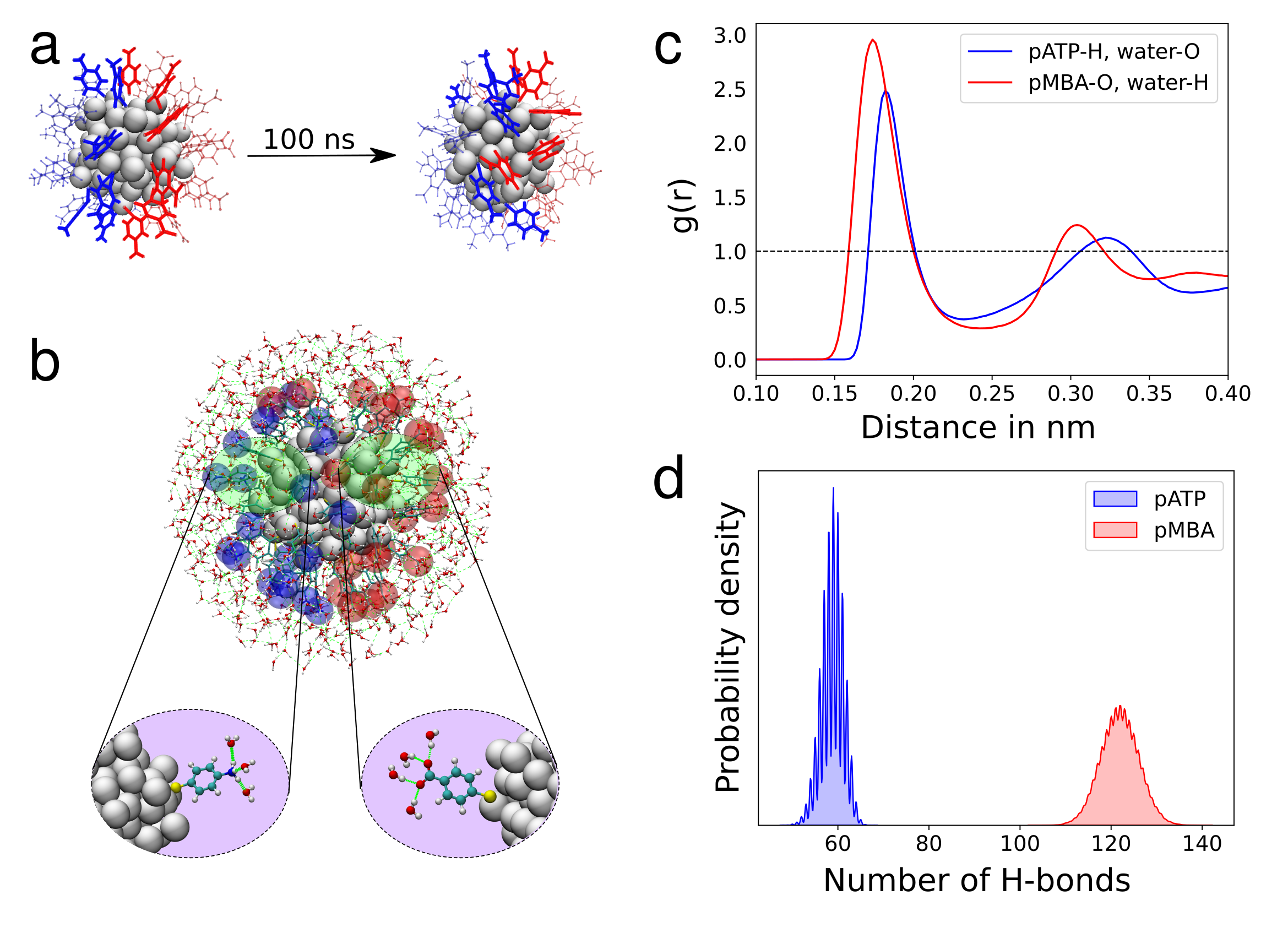}
    \caption{Analyzing the Au-102 J-MPC monomer. (a)The primarily even interface changes into the 'crooked' interface.(b) Au-102 J-MPC surrounded by water(solvent). The orientation of the positively and negatively charged ligands at the Janus boundary region.  (c) Radial Distribution Functions. (d) Hydrogen bond distribution.}
    \label{fig:mono}
\end{figure}

The zwitterionic J-MPC, Au$_{102}$(S-\textit{p}C$_6$H$_4$COO$^-$)$_{22}$(S-\textit{p}C$_6$H$_4$NH$_3^+$)$_{22}$ we have modeled is not primarily symmetrical. A `crooked' interface is the result of the electrostatic and ion-dipole interactions between the oppositely charged \textit{p}MBA and \textit{p}ATP ligands at the Janus boundary interface (Fig \ref{fig:mono}a). Furthermore, on analyzing Radial Distribution Functions (RDF), we observe higher correlation for the first coordination shell of the two oxygen atoms of the carboxylic (COO$^-$) group of \textit{p}MBA with the corresponding hydrogen atoms of the water molecules as opposed to the three hydrogen atoms of NH$^{3+}$ of \textit{p}ATP with the oxygen atom of the water molecules as depicted in Fig \ref{fig:mono}b and c. This suggests stronger interactions between the water molecules and the \textit{p}MBA ligands as compared to the \textit{p}ATP ligands. The contrast in the number of hydrogen bonds (Fig \ref{fig:mono}d) on either side of the hemisphere further manifests the asymmetric solvation of the J-MPC (Fig. \ref{fig:mono}c and d). The ratio between the H's of H$_2$O and the O's of COO$^-$ with respect to the ratio between the O of water and the H's of NH$^{3+}$ serves as the differentiating parameter. 
 
\subsection{Dimer simulations} 

Dimerization is the pivotal elementary step in the process of molecular aggregation and delineates the characteristics of the self-assembly process. To understand the dimerization process, we simulated a system containing two equilibrated Au$_{102}$(\textit{p}MBA)$_{22}$(\textit{p}ATP)$_{22}$ monomers placed randomly in a large cubic simulation box of edge length ~12 nm filled with water molecules. The system, post-minimization and equilibration, was simulated for 100 ns in the NPT ensemble. Within a short simulation time $\sim$15 ns, the two monomers, strongly attracted by electrostatic interactions, quickly formed a dimer with a COM-COM distance of $\sim$2.4-2.5 nm. This dimeric state remained stable throughout the entire simulation time. Due to the strong electrostatic interactions between the oppositely charged surface ligands, the J-MPCs orient themselves favoring the formation of a stable dimer.

\begin{figure}[!htbp]
    \centering
    \includegraphics[width=165mm]{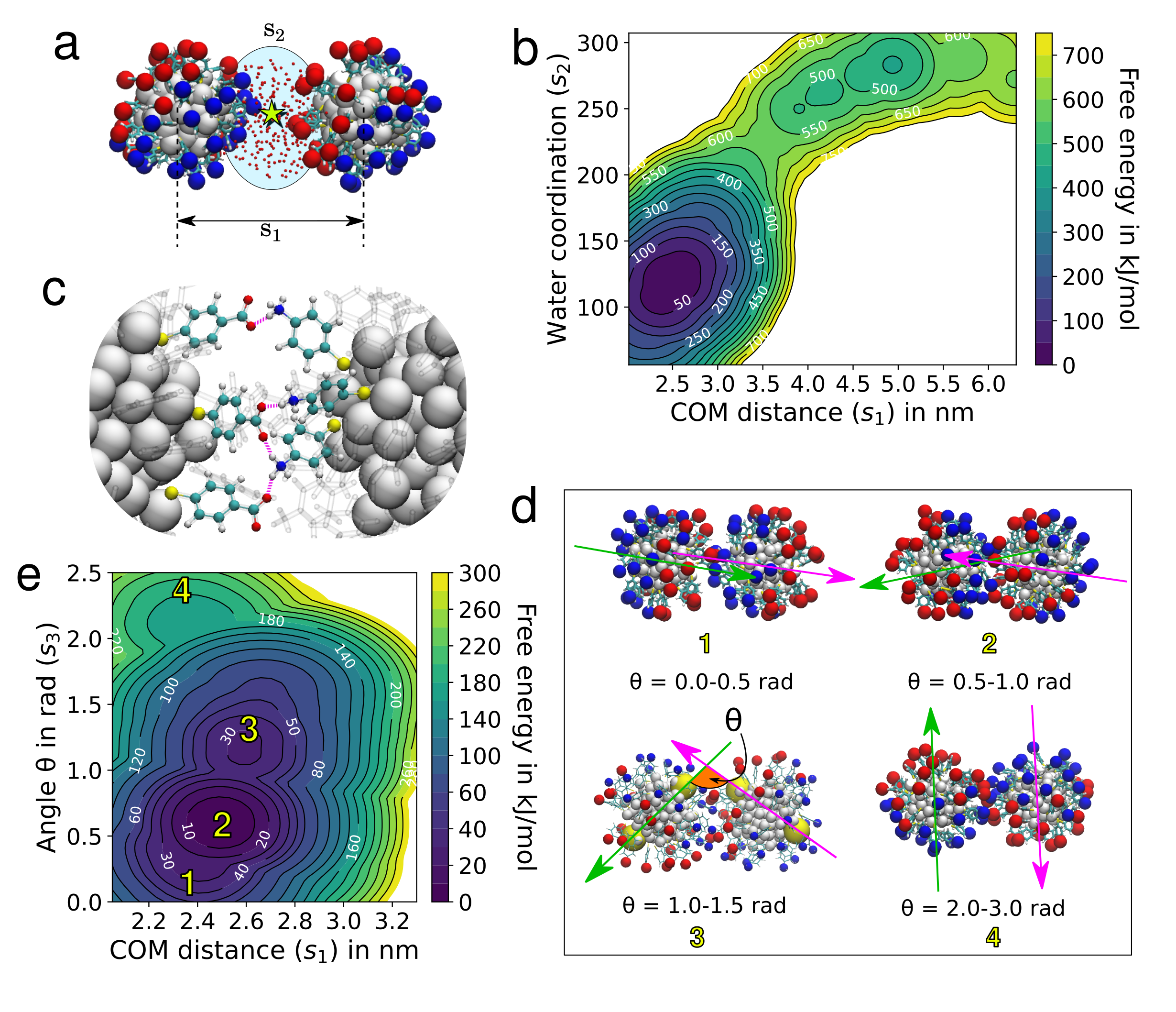}
    \caption{Dimerization of J-MPCs. (a)CV representation; s$_1$ is the COM-COM distance between the two monomers, and s$_2$ is the number of coordinated solvent molecules at the interface of the dimer. (b) Free energy surface (FES) as a function of s$_1$ and s$_2$. (c) Snapshot showing inter-MPC ligands interactions. (d) Snapshot of J-MPC dimers (1-4) distinguished based on $\theta$ value. Angle $\theta$ (s$_3$) used to characterize dimers based on orientation was chosen between two vectors (represented in green and magenta colors) passing through the selected sulfur atoms (represented in yellow color) present at the center of each hemisphere of the J-MPCs. (e) Free energy surface (FES) as a function of s$_1$ and s$_3$.}
    \label{fig:dimer}
\end{figure}

To understand the mechanism and thermodynamics of the dimerization process, we carried out On-the-fly probability-based enhanced sampling (OPES$_e$, the explore variant)\cite{Invernizzi2020} simulations that efficiently sample the monomer to dimer transformations and help us to calculate the underlying free energy surface (FES) of dimerization. An advancement of the metadynamics approach, OPES periodically deposits bias potentials in the form of Gaussians as a function of selected degrees of freedom, called collective variables (CVs), and calculates on-the-fly, the equilibrium probability distribution. Accurately sampling of the phase space depends on the CV's quality in distinguishing between various metastable states. In the case of dimerization, we have chosen two CVs, namely, s$_1$ and s$_2$. The s$_1$ represents the COM-COM distance between the Au-core of the MPCs, whereas, s$_2$ describes the number of solvent molecules at their interface (Fig. \ref{fig:dimer}a). From our previous experience \cite{Tiwari2023}, we realized that the solvent CV, s$_2$ is beneficial for efficient sampling of the monomer-dimer transitions. Using these two CVs an OPES$_e$ simulation was performed for $\sim$2 $\mu$s, and multiple back-and-forth transitions between the two metastable states were observed. (Fig Sxx)

The free energy surface plot in Fig \ref{fig:dimer}b obtained by OPES$_e$ simulation, maps the relative stability of the monomeric and dimeric states. From the FES, the dimeric state is found to have very high stability. It is also not surprising that the dimerization-free energy barrier is quite high; this is due to strong electrostatic interactions between the oppositely charged ligands of the MPCs at the dimer interface  (Fig \ref{fig:dimer}c). The high free energy barrier for the detachment of the dimer to form monomers indicates that the dimerization process is irreversible.\cite{Gentili2022rev} Interestingly, the strong electrostatic interactions between the monomers result in the formation of directional self-assembled structures, which we will discuss in the subsequent section.

In the OPES simulation, we observed that the zwitterionic J-MPCs interact with one another in various dipolar orientations that correspond to several microstates in the dimeric free energy basin (see Fig. \ref{fig:dimer}d). To quantify the dipolar orientation in the dimeric form, we chose an angle ($\theta$) between two vectors passing through the sulfur atoms present at the center of each hemisphere of the J-MPC molecule as depicted in structure-3 of Fig. \ref{fig:dimer}d. Further, we calculated the free energy surface using s$_1$ and s$_3$ ($\theta$) with the previously run OPES$_e$ simulation (Fig. \ref{fig:dimer}e). The dimers were categorized into three groups \textit{viz.}, parallel ($\theta<$ 0.5), twisted (0.5 $>\theta>$ 2), and anti-parallel ($\theta>$ 2). The representative dimeric structures from these groups have been shown in Fig. \ref{fig:dimer}d. Twisted dimers are the most frequently observed because of two main reasons (i) twisting increases the surface area of contact between the oppositely charged ligands leading to strong electrostatic attraction, and (ii) higher statistical probability in comparison to the parallel/anti-parallel dimers. The twisted dimers show two basins (2 \& 3) separated by a small energy barrier of $\sim$35 kJ/mol. (Fig. \ref{fig:dimer}e) Most of the dimeric states observed lie between 2.4-2.8 nm COM-COM distance accompanied by strong electrostatic interaction between the oppositely charged ligands. Whereas more compact (s$_1<$ 2.4 nm) dimeric states were less stable due to reduced electrostatic interactions of the ligands as a result of MPCs' close proximity.

\subsection{Self-assembly} 

\begin{figure}[!htbp]
    \centering
    \includegraphics[width=160mm]{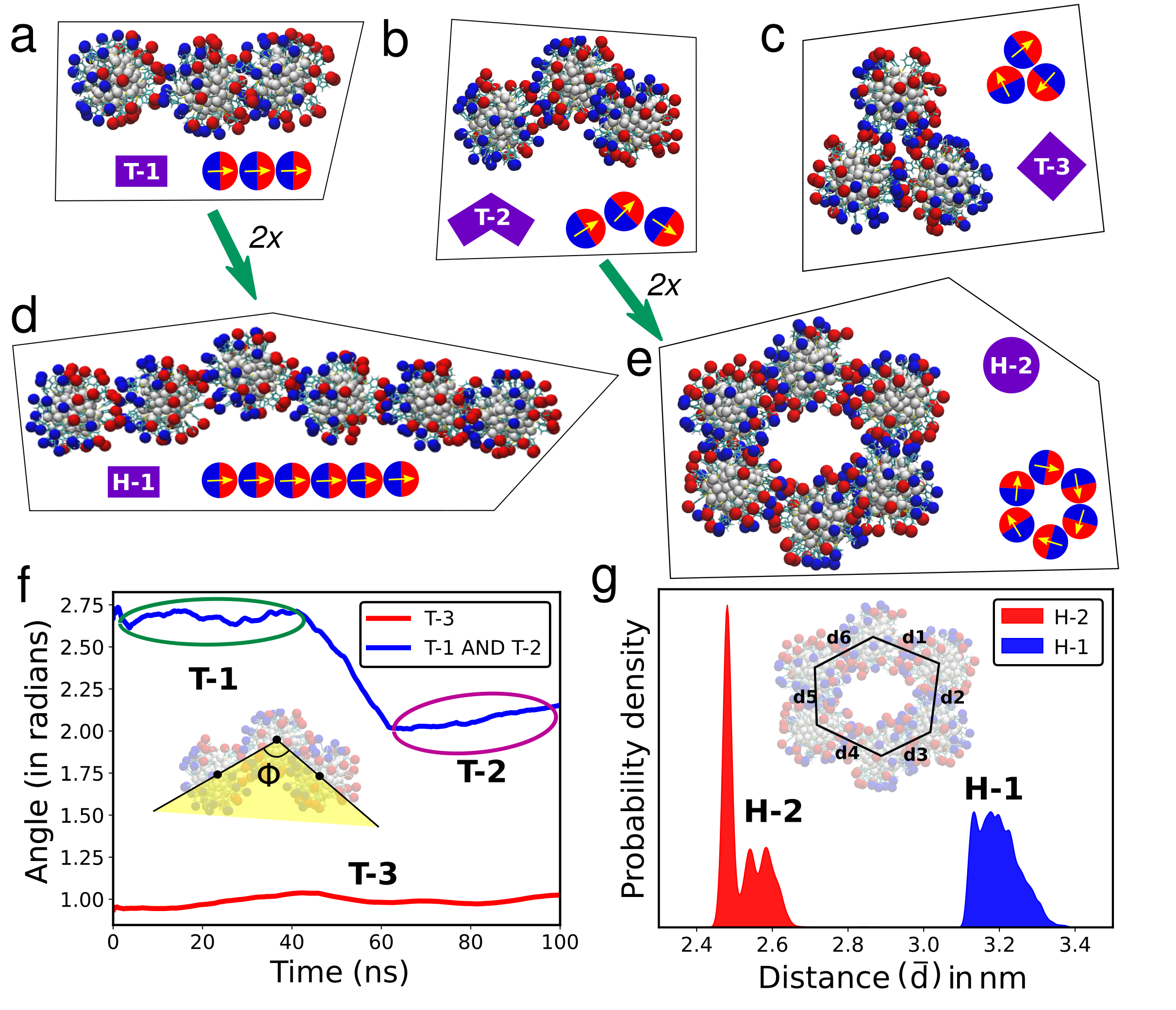}
    \caption{Self-assembly of J-MPCs. (a) Linear trimer, T-1. (b) Bent trimer, T-2. (c) Triangular trimer, T-3. (d) Preformed linear hexamer, H-1 (e) distances in the preformed cyclic hexamer, H-2, the average COM-COM distance, $\bar{d} = \sum_{i}^{6}d_i/6$ (f) Plot representing the variation in angles with time for the 3 trimers. (g) Plot representing the average COM-COM distance between each MPC center of hexamers.}
    \label{fig:tri-hex}
\end{figure}

Next, we focused on studying the formation of multimeric structures of J-MPCs. Here we explored various possibilities of self-assembled multimers by running multiple independent simulations of systems containing varying numbers of J-MPCs. Firstly we focused on the formation of trimeric structures. A simulation was set up by closely placing a monomer linearly to the dimer, and we checked for the probability of the formation of a linear trimer (T-1) (Fig. \ref{fig:tri-hex}a). The MPCs do arrange and aggregate linearly but after $\sim$40ns the angle ($\omega$) between the COMs of each MPC decreased to form the bent trimer (T-2) (Fig. \ref{fig:tri-hex}b) which retains its shape all through the entirety of the 100 ns simulation. Further, an independent simulation was performed by placing an equilibrated monomer angularly near the equilibrated dimer which led to the formation of triangular trimer T-3 (Fig. \ref{fig:tri-hex}c). Later the hexameric structures were formed using linear (T-1) and bent (T-2) trimers. Placing two T-1 linearly led to the formation of a linear hexamer (H-1) (Fig. \ref{fig:tri-hex}d), whereas, when two T-2 were placed facing each other, the cyclic hexamer (H-2) was formed (Fig. \ref{fig:tri-hex}e). All three trimers were distinguished based on the angle $\omega$ present between the COMs of the J-MPC molecules (Fig. \ref{fig:tri-hex}f) whereas, two hexameric structures were distinguished based on the average distance ($\bar{d}$ between the consecutive J-MPC molecules (Fig. \ref{fig:tri-hex}g).

\begin{figure}[!ht]
    \centering
    \includegraphics[width=165mm]{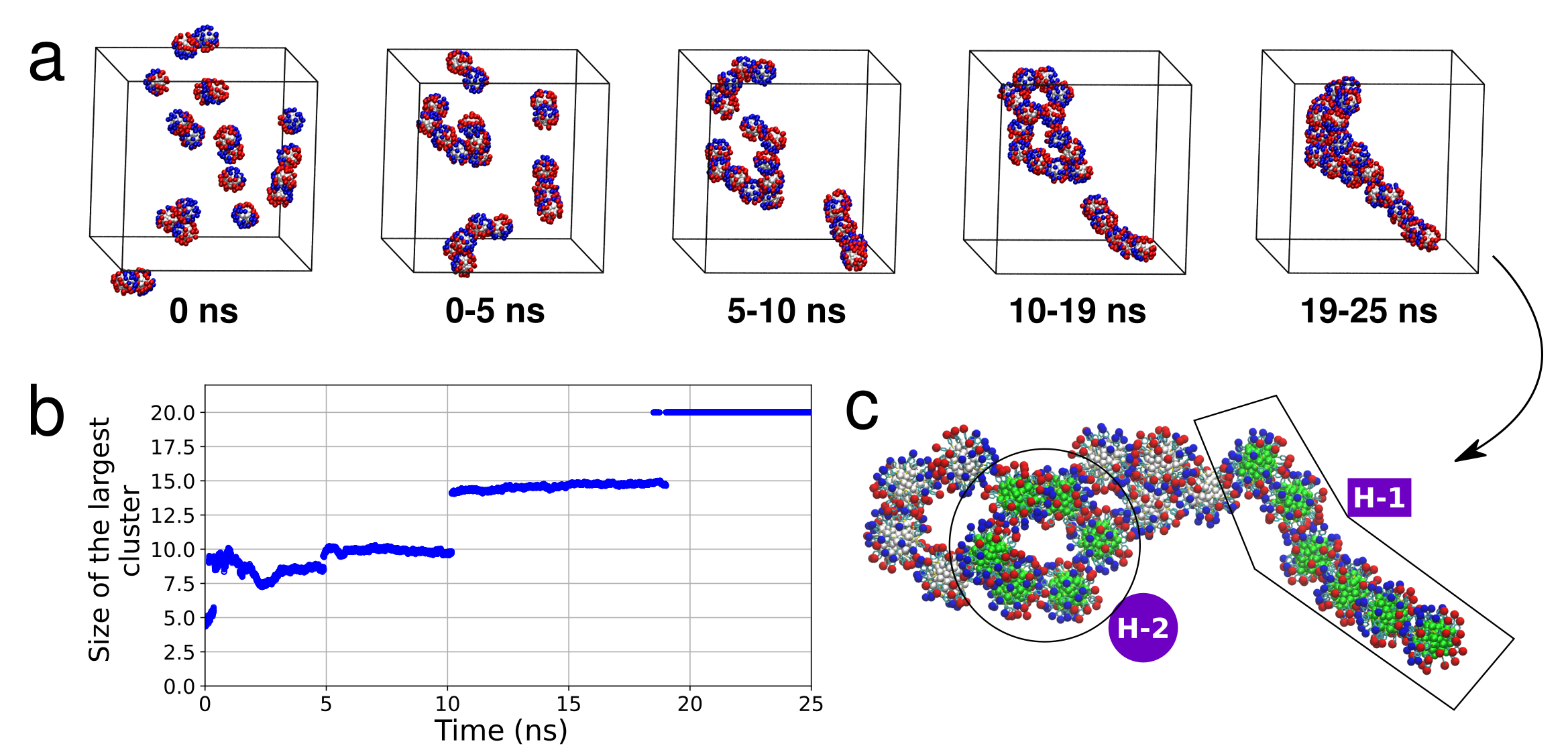}
    \caption{Analyzing the 20 randomly dispersed J-MPC system. (a) The evolution of the fully clustered multimer with time. (b) Using cluster size as a function of simulation time to characterize the self-assembly. (c) The emergence of the 2 pre-formed hexamers H-1 and H-2}
    \label{fig:multimer}
\end{figure}

In the previous section, we have presented the results of the simulations of pre-formed linear and ring-hexamer clusters. These were based on our anticipation that J-MPCs, owing to their bipolar nature, would form such structural patterns. To verify our anticipation, we subsequently carried out a series of simulations with randomly dispersed MPCs in the solution and monitored their self-assembly process. We have simulated two sets - in the first, six monomers were kept hoping that they would aggregate to form the hexameric ring, and in the other, twenty monomers were placed to observe the evolution of various self-assembled patterns. Interestingly, in set 1, the randomly dispersed six J-MPC monomers aggregated to form both linear and cyclic hexamers (Fig Sxx). On the other hand, the larger system containing twenty MPCs showed the formation of multiple clusters with varying sizes that further aggregated to form a large cluster containing both cyclic hexameric units and linear chains (see Fig. \ref{fig:multimer}c).

\section{CONCLUSIONS}
Our work on the self-assembly of MPCs shows that intriguing hierarchical structures can be assembled by the rational design of ligand functional groups and tuning the ligand charge states. The J-MPC monomer exhibits asymmetric solvation, with the pMBA ligands binding more strongly to the solvent in comparison to the pATP ligands. The oppositely charged ligands exert electrostatic interactions resulting in the formation of a crooked Janus boundary interface. Strong electrostatic interactions between the monomers make the dimerization an irreversible process. Further, the overall molecular dipole of the monomers helps the formation of directional linear and hexameric-ring-like superstructures. Our work showcases the potential of rational ligand functionalization and the ligand charge-state mediated MPCs self-assembly. The day when these J-MPCs are synthesized seems imminent, with uses spanning beyond drug delivery, biomedical imaging, cancer therapy, and catalysis.

\vspace{20pt}
\noindent
{\bf Corresponding Author*}\\
E-mail: tkarmakar@chemistry.iitd.ac.in\\

\noindent
{\bf Author Contributions}\\
$^\dag$ Authors have contributed equally.\\

\noindent
{\bf ORCID:}\\
Vikas Tiwari: 0000-0003-4510-253X\\
Tarak Karmakar: 0000-0002-8721-6247\\

\noindent
{\bf Present Address}\\
Department of Chemistry\\
Indian Institute of Technology, Delhi \\
Hauz Khas, New Delhi 110016\\
Delhi, India\\

\noindent
{\bf Notes:}\\ 
The authors declare no competing financial interest.\\

%\newpage
%{\bf 7. ACKNOWLEDGEMENTS}\\
\begin{acknowledgement}
    A.B. acknowledges IIT Delhi for the MTech fellowship. V.T. thanks the Ministry of Education, Govt. of India for the PMR Fellowship.  T.K. acknowledges the Science and Engineering Research Board (SERB), New Delhi, India for the Start-up Research Grant (File No. SRG/2022/000969). We also acknowledge IIT Delhi for the Seed Grant. We thank the IIT Delhi HPC facility for computational resources. 
\end{acknowledgement}

%\newpage
\bibliography{ref}{}

\iffalse
\newpage
\vspace{30pt}
\begin{figure}%[!ht]
  \begin{center}
    \includegraphics[width=1.0\textwidth]{TOC.png}
  \end{center}
  {\bf Table of Contents}
\end{figure}
\fi 

\end{document}